\documentclass[submission,copyright,creativecommons]{eptcs}
\providecommand{\event}{FROM 2019} 
\usepackage{underscore}           

\usepackage{graphicx}
\usepackage{latexsym}

\usepackage{amsthm}
\usepackage{epsfig}
\usepackage{subcaption}
\usepackage{amssymb}
\usepackage{amstext}
\usepackage{amsmath}
\usepackage{amsthm}

\newcommand{\bean}{\begin{eqnarray*}}
	\newcommand{\eean}{\end{eqnarray*}}
\newcommand{\bp}{\begin{proof}}
	\newcommand{\ep}{\end{proof}}

\def\.{\!\!\!.}

\newcommand{\lra}{\longrightarrow}

\newcommand{\y}{\,\wedge\,}

\title{Networks of Uniform Splicing Processors\thanks{Work supported by a grant of the Romanian National Authority for Scientific Research and Innovation, project number POC P-37-257.}}
\author{Victor Mitrana
\institute{National Institute for Research and Development of Biological Sciences,\\
	Independentei Bd. 296, Bucharest, Romania}
\email{victor.mitrana@upm.es}
\and
Mihaela P\u aun
\institute{National Institute for Research and Development of Biological Sciences,\\
	Independentei Bd. 296, Bucharest, Romania}
\email{mihaela.paun@gmail.com}
\and
Jose Angel Sanchez Martin\qquad\qquad Jose Ramon Sanchez Couso
\institute{Department of Information Systems, Polytechnic University of Madrid, \\
	Crta. de Valencia km. 7 - 28031 Madrid}
\email{\quad joseangel.sanchez.martin@alumnos.upm.es \quad\qquad jcouso@eui.upm.es}
}
\def\titlerunning{Networks of Uniform Splicing Processors}
\def\authorrunning{V. Mitrana, M. P\u aun, J.A. Sanchez Martin \& J.R. Sanchez Couso}
\begin{document}
\maketitle

\begin{abstract}
In this note we consider a new variant of network of splicing processors which simplifies the general model such that filters remain associated with nodes but the input and output filters of every node coincide. This variant is called {\it network of uniform splicing processors}. Although the 
communication in the new variant seems less powerful, being based on simpler filters, the new variant is sufficiently powerful to be computationally complete. The main result is that nondeterministic Turing machines can be simulated 
by networks of uniform splicing processors. Furthermore, the simulation is time efficient.
\end{abstract}

\section{Introduction}

Computational models inspired by different biological phenomena turned out to be theoretically able to
efficiently solve intractable problems. The main computational features of these models are abstracted from the way in which nature evolves.
These computational models have appeared in the last two decades and have been vividly investigated from a formal point of view. 
For a survey of several classes of bio-inspired computational models the reader is referred to \cite{handbookNC}.

Along these lines, networks of bio-inspired processors form a class of highly parallel and distributed computing models inspired and abstracted from different biological phenomena.
Networks of bio-inspired processors resemble other models of computation with similar or different origins:
{\it evolutionary systems} inspired by the evolution of
cell populations \cite{erzsi}, {\it tissue-like P systems} \cite{tissue} in the membrane computing area \cite{membrane}, {\it networks of parallel language processors} as a formal languages generating device \cite{arto}, {\it flow-based programming} as a well-known programming paradigm \cite{Morrison},
{\it distributed computing using mobile programs} \cite{Gray},
{\it Connection Machine}, viewed as a network of microprocessors processing one bit per unit time in the shape of a hypercube \cite{connection}, etc.
Networks of bio-inspired processors may be informally described as a graph whose vertices are processors running operations on data structured as strings, pictures, multisets.
Two main types of string processors have been considered so far: evolutionary processors and splicing processors.

A splicing processor \cite{Manea2007} performs an operation called {\it splicing} that is inspired from the recombination of DNA molecules under the effect of different types of enzymes \cite{Head-Handbook}.
This phenomenon, called {\it splicing}, allows to genetically modify a biological entity for different purposes like: more resistant plants, organisms better adapted to weather changes, production of hormones, etc.
The chemicals involved in the recombination of DNA sequences are two types of enzymes:
restriction enzymes which cut the DNA at specific sites (called recognition sites) yielding two fragments with the so-called ``sticky ends'', and ligase which rejoin fragments with sticky ends.
A computational model based on an operation abstracted from the splicing operation described above has been defined in \cite{Head87}. The model viewed as a language generating device is called {\it splicing system}.
Roughly speaking, the two DNA molecules are represented by strings while the restriction enzymes are represented by quadruples of strings, called splicing rules, indicating the sites where the two strings are to be cut. The compatibility for rejoining is defined by the fact that two fragments can be rejoined if they were obtained by applying the same splicing rule.

Networks of splicing processors (NSP), were introduced in \cite{Manea2007}.
The NSP model resembles some features of the
test tube distributed systems based on splicing introduced in
\cite{dtts} and further investigated in \cite{paun}. The
differences between the models considered in \cite{dtts} and \cite{Manea2007} are precisely described in \cite{Manea2007}.
In \cite{Manea2007} one also mentions the differences between NSP and the time-varying
distributed H systems, another generative model based on splicing introduced in \cite{tvd}.

The computation in
a network of splicing processors consists of a sequence of steps,
splicing and communication, which alternate with each other until a predefined
condition is satisfied.
In each splicing step, all processors simultaneously apply their rules on the data
existing in the nodes hosting them. Furthermore, the data in each node is organized in the form of multisets
of strings (each string may appear in an arbitrarily large number of copies),
and all copies are processed in parallel so that all the possible
events that can take place do actually take place.
In each communication step, two actions are done according to different strategies:

(i) all the nodes simultaneously send out the data they contain after a splicing step to all adjacent nodes;

(ii) all the nodes simultaneously handle all the arriving data.\\
The communication strategies considered so far are based on filters that allow or forbid
strings to enter nodes or go out from nodes. These filters are defined mainly by two types of conditions: syntactical conditions (random-context conditions, membership to regular languages, semi-conditional conditions) and semantic conditions, where polarization is just a very simple case.
In the case of filters based on syntactical conditions in networks of splicing processors, there are two variants: (i) each node has an input and an output filter that could be different \cite{Manea2007} and (ii) 
the filters as in the previous case of two adjacent nodes collapse on the edge between them such that each edge, which acts as a bidirectional channel between the nodes, has a unique filter \cite{jucs}.

We consider here a new variant, somehow ''in between'' the two aforementioned variants. More precisely, the two filters associated with nodes collapse to only one such that the input and the output filters coincide.

\section{Basic definitions}

We assume the reader is familiar with the basic notions of the formal language theory. In the sequel, we summarize the main concepts and notations used in this work;
for all unexplained notions the reader is referred to \cite{handbook}.

An {\it alphabet} is a finite and nonempty set of symbols. The
cardinality of a finite set $A$ is written $card(A)$. Any finite
sequence of symbols from an alphabet $V$ is called a {\it string} over
$V$. The set of all strings over $V$ is denoted by $V^*$ and the empty
string is denoted by $\lambda$. The length of a string $x$ is denoted by
$|x|$ while $alph(x)$ denotes the minimal alphabet $W$ such that
$x\in W^*$. 

We continue with the formal definition of the splicing operation following \cite{Head-Handbook}. A \emph{splicing rule} over a finite alphabet $V$ is a quadruple of strings of the
form $[(u_{1},u_{2});(v_{1},v_{2})]$ such that $u_{1}$, $u_{2}$,
$v_{1}$, and $v_{2}$ are in $V^{*}$.
For a splicing rule $r = [(u_{1},u_{2});(v_{1},v_{2})]$ and for
$x,y,z\in V^{*}$, we say that $r$ produces $z$ from $x$ and $y$
(denoted by $(x,y)\vdash_{r}z$) if there exist some $x_{1},
x_{2}, y_{1}, y_{2}\in V^{*}$ such that $x=x_{1}u_{1}u_{2}x_{2}$,
$y=y_{1}v_{1}v_{2}y_{2}$, and $z=x_{1}u_{1}v_{2}y_2$.
For a language $L$ over $V$ and a set of splicing rules $R$ we
define\\
\centerline{$\sigma_{R}(L)=\{z\in V^{*}\mid \exists u, v \in L, \exists r \in R\mbox{ such that }
(u, v) \vdash_r z\}.$}

For two disjoint and nonempty subsets $P$ and $F$ of an alphabet $V$ and
a string $z$ over $V$, we define the predicates
\begin{center}
	\begin{tabular}{llll}
		$\varphi^{(s)}(z;P,F)\equiv$ & $P\subseteq alph(z)$ &$\y$ & $F\cap
		alph(z)=\emptyset$\\
		$\varphi^{(w)}(z;P,F)\equiv$ & $alph(z)\cap P \ne \emptyset$ & $\y$ & $F\cap alph(z)=\emptyset $.
	\end{tabular}
\end{center}

The construction of these predicates is based on {\it random-context
	conditions} defined by the two sets $P$ ({\it permitting contexts/symbols})
and $F$ ({\it forbidding contexts/symbols}). Informally, the former condition requires that all permitting 
symbols are and no forbidding symbol is present in $w$,  while the latter 
is a weaker variant such that at least one permitting symbol appears in $w$ but 
still no forbidding symbol is present in $w$.  

For every language $L\subseteq V^*$ and $\beta\in \{(s),(w)\}$, we define:\\
\centerline{$\varphi^\beta(L,P,F)=\{w\in L\mid \varphi^\beta(w;P,F)\}.$}

A \emph{splicing processor} over $V$ is a $6$-tuple $(S,A,PI,FI,PO,FO)$,
where:\\
-- $S$ is a finite set of splicing rules over $V$.\\
-- $A$ is a finite set of {\it auxiliary strings} over $V$. These auxiliary 
strings are to be used, together with the existing strings, in the splicing steps of the
processors. Auxiliary strings are available at any moment.\\
-- $PI,FI\subseteq V$ are the {\it input} permitting/forbidding
contexts of the processor, while $PO,FO\subseteq V$ are the {\it output}
permitting/forbidding contexts of the processor (with $PI \cap FI=
\emptyset$ and $PO \cap FO=\emptyset$).

A splicing processor as above is said to be {\it uniform} if $PI=PO=P$ and $FI=FO=F$. For the rest
of this note we deal with uniform splicing processors only. 
We denote the set of uniform splicing processors over $V$ by $USP_V$.

A \emph{network of uniform splicing processors} (NUSP for
short) is a $9$-tuple $\Gamma=(V,U,<,>,G,\mathcal{N},\alpha,\underline{In}$, $\underline{Halt})$,
where:\\
	$\bullet$ $V$ and $U$ are the input and network alphabet, respectively, $V\subseteq U$, and, also, $<,>\in U\setminus V$ are two special symbols.\\ 
	$\bullet$ $G=(X_G,E_G)$ is an undirected graph without loops with the set of nodes
	$X_G$ and the set of edges $E_G$. Each edge is given in the form of a
	binary set. $G$ is called the {\it underlying
		graph} of the network.\\
	$\bullet$ $\mathcal{N}:X_G\lra USP_U$ is a mapping which associates with each
	node $x\in X_G$ the
	splicing processor $\mathcal{N}(x)=(S_x,A_x,P_x,F_x)$.\\
	$\bullet$ $\alpha: X_G\lra \{(s),(w)\}$ defines the type of the
	filters of a node. \\
	$\bullet$ $\underline{In}, \underline{Halt} \in X_G$  are the {\it input} and 
	the {\it halting} node of $\Gamma$, respectively.\\

The \emph{size} of $\Gamma$ corresponds to the number of nodes in
the graph, i.e. $card(X_G)$. A \emph{configuration} of an NUSP
$\Gamma$ is a mapping $C:X_G\rightarrow 2^{U^*}$ which associates
a set of strings with every node of the graph. Although a configuration is a 
multiset of strings, each one appearing in an arbitrary number of copies, for sake 
of simplicity we work with the support of this multiset. A configuration can be
seen as the sets of strings, except the auxiliray ones, which are present in any node at a given
moment. For a string $w\in V^*$ the initial configuration of
$\Gamma$ on $w$ is defined by $C^{(w)}_0(\underline{In}) = \{\langle w
\rangle\}$ and $C^{(w)}_0 (x) =\emptyset$ for all other $x\in
X_G$. 

There are two ways to change a configuration, by a splicing step
or by a communication step. When changing by a splicing step, each
component $C(x)$ of the configuration $C$ is changed according to
the set of splicing rules $S_x$, whereby the strings in the set
$A_x$ are available for splicing. Formally, configuration $C'$ is
obtained in one splicing step from the configuration $C$, written
as $C\Rightarrow C'$, iff for all $x\in X_G$, the following holds:\\
\centerline{$C'(x)=\sigma_{S_x}(C(x)\cup A_x).$}\\
In a communication step, each processor $x$ sends out all strings that
can pass its filter. They are received by all the other nodes $y$
in the graph, connected to $x$, provided that they pass the filter of $y$. Note that,
according to this definition, strings that can leave a node are
sent out even if they cannot pass the filter of any node. In this case
we will say that they are lost. Formally, $C'$ is obtained from
$C$ (we write $C'\models C)$ iff for all $x\in X_G$\\
\centerline{$C'(x)=(C(x)-\varphi^{\beta(x)}(C(x),P_x,F_x))\cup\displaystyle{\bigcup_{\{x,y\}\in
	E_G}(\varphi^{\beta(y)}(C(y),P_y,F_y)\cap\varphi^{\beta(x)}(C(y),P_x,F_x))}$}\\
holds.
For an NUSP $\Gamma$, the computation on an input string $w$ is a
sequence of configurations $C^{(w)}_0$, $C^{(w)}_1$, $C^{(w)}_2,...$,
where $C^{(w)}_0$ is the initial configuration of $\Gamma$ on $w$,
$C^{(w)}_{2i}\Rightarrow C^{(w)}_{2i+1}$ and
$C^{(w)}_{2i+1}\models C^{(w)}_{2i+2}$, for all $i\ge 0$. A
computation \emph{halts} if there exist a configuration in which the set of strings existing in the halting
	node $\underline{Halt}$ is non-empty. This is an \emph{accepting computation}.
The language accepted by $\Gamma$
is defined as 
\begin{center}$L(\Gamma)=\{w\in V^*\mid \Gamma's$
	computation on $w$ is an accepting computation$\}$.
\end{center} 

We define two computational complexity measures using NUSP as the
computing model.  To this aim we consider an NUSP $\Gamma$ with the input alphabet $V$
that halts on every input. The {\it time complexity} of the
finite computation $C_0^{(x)}$, $C_1^{(x)}$, $C_2^{(x)}$, $\dots
C_m^{(x)}$ of $\Gamma$ on $x\in V^*$ is denoted by $Time_{\Gamma}(x)$ and
equals $m$.
The time complexity of $\Gamma$ is the partial function
from {\bf N} to {\bf N},\\
\centerline{$Time_{\Gamma}(n)=\mbox{max}\{Time_{\Gamma}(x)\mid x\in
V^*, |x|=n\}.$}

\section{Main result}

The main result of this note is a time efficient simulation of Turing machines by NUSP.
Formally,

\noindent{\bf Theorem.}{\it \\
	1. If a language is accepted by a nondeterministic Turing machine, then it is accepted by an NUSP.\\
	2. If a language is accepted by a nondeterministic Turing machine in $\mathcal{O}(f(n))$ time, then it is decided by an NUSP in time  $\mathcal{O}(f(n))$.}

\noindent{\bf Sketch of the proof.} The network contains the input and halting nodes, two nodes $\underline{Sim}$ and $\underline{Res}$, as well as 2 further nodes for each transition of the Turing machine. We shall not give the formal descriptions of the nodes (sets of rules, axioms, permitting and forbidding symbols, respectively), but we informally describe how the network works. The input string $<w>$ is 
transformed into $<^{q_0} wB\$>'$ by two splicing steps in $\underline{In}$, where $q_0$ is the initial state of the Turing machine and $B$ is the blank symbol. Now the network performs a ``rotate-and-simulate'' strategy.
The obtained string enters $\underline{Sim}$, where the first symbol $<^{q_0}$ (inductively,
$<^q$) is simultaneously replaced by $<^{q_0,a,s,b,R}$ or $<^{q_0a,s,b,L}$, where $(q_0,a,s,b,R)$ and $(q_0,a,s,b,L)$ are transitions, in different copies of $<^{q_0} wB\$>'$. Now, the new strings enter the nodes associated with the corresponding transitions 
where that transition is simulated. Each transition is simulated by a constant number of splicing. Then all the strings enter node $\underline{Res}$, where after two splicing steps, either the whole process described above starting in $\underline{Sim}$ is resumed, or a string enters $\underline{Halt}$ and the computation halts.
\hspace*{\fill} $\Box$

\nocite{*}
\bibliographystyle{eptcs}
\bibliography{from}

\begin{thebibliography}{10}
\providecommand{\bibitemdeclare}[2]{}
\providecommand{\surnamestart}{}
\providecommand{\surnameend}{}
\providecommand{\urlprefix}{Available at }
\providecommand{\url}[1]{\texttt{#1}}
\providecommand{\href}[2]{\texttt{#2}}
\providecommand{\urlalt}[2]{\href{#1}{#2}}
\providecommand{\doi}[1]{doi:\urlalt{http://dx.doi.org/#1}{#1}}
\providecommand{\bibinfo}[2]{#2}

\bibitemdeclare{article}{dtts}
\bibitem{dtts}
\bibinfo{author}{E.~\surnamestart Csuhaj-Varj{\'u}\surnameend},
  \bibinfo{author}{L.~\surnamestart Kari\surnameend} \&
  \bibinfo{author}{G.~\surnamestart P\u{a}un\surnameend}
  (\bibinfo{year}{1996}): \emph{\bibinfo{title}{Test tube distributed systems
  based on splicing.}}
\newblock {\sl \bibinfo{journal}{Computers and Artificial Intelligence}}
  \bibinfo{volume}{15}, pp. \bibinfo{pages}{211--232}.

\bibitemdeclare{article}{erzsi}
\bibitem{erzsi}
\bibinfo{author}{E.~\surnamestart Csuhaj-Varj{\'u}\surnameend} \&
  \bibinfo{author}{V.~\surnamestart Mitrana\surnameend} (\bibinfo{year}{2000}):
  \emph{\bibinfo{title}{Evolutionary systems: A language generating device
  inspired by evolving communities of cells.}}
\newblock {\sl \bibinfo{journal}{Acta Informatica}} \bibinfo{volume}{36}, pp.
  \bibinfo{pages}{913--926}, \doi{https://doi.org/10.1007/s002360050178}.
\newblock
  \urlprefix\url{https://link.springer.com/article/10.1007/s002360050178}.

\bibitemdeclare{article}{arto}
\bibitem{arto}
\bibinfo{author}{E.~\surnamestart Csuhaj-Varj{\'u}\surnameend} \&
  \bibinfo{author}{A.~\surnamestart Salomaa\surnameend} (\bibinfo{year}{2005}):
  \emph{\bibinfo{title}{Networks of parallel language processors.}}
\newblock {\sl \bibinfo{journal}{Lecture Notes in Computer Science (LNCS)}}
  \bibinfo{volume}{1218}, pp. \bibinfo{pages}{299--318},
  \doi{https://doi.org/10.1007/3-540-62844-4_22}.
\newblock
  \urlprefix\url{https://link.springer.com/chapter/10.10072F3-540-62844-4_22}.

\bibitemdeclare{article}{jucs}
\bibitem{jucs}
\bibinfo{author}{C.~\surnamestart Dr\u{a}goi\surnameend},
  \bibinfo{author}{F.~\surnamestart Manea\surnameend} \&
  \bibinfo{author}{V.~\surnamestart Mitrana\surnameend} (\bibinfo{year}{2007}):
  \emph{\bibinfo{title}{Accepting networks of evolutionary processors with
  filtered connections.}}
\newblock {\sl \bibinfo{journal}{Journal of Universal Computer Science}}
  \bibinfo{volume}{13}, pp. \bibinfo{pages}{1598--1614},
  \doi{https://doi.org/10.3217/jucs-013-11-1598}.
\newblock
  \urlprefix\url{http://www.jucs.org/jucs_13_11/accepting_networks_of_evolutionary}.

\bibitemdeclare{article}{Gray}
\bibitem{Gray}
\bibinfo{author}{R.~\surnamestart Gray\surnameend},
  \bibinfo{author}{D.~\surnamestart Kotz\surnameend},
  \bibinfo{author}{S.~\surnamestart Nog\surnameend},
  \bibinfo{author}{D.~\surnamestart Rus\surnameend} \&
  \bibinfo{author}{G.~\surnamestart Cybenko\surnameend} (\bibinfo{year}{1997}):
  \emph{\bibinfo{title}{Mobile agents: The next generation in distributed
  computing.}}
\newblock {\sl \bibinfo{journal}{Proceedings of IEEE International Symposium on
  Parallel Algorithms Architecture Synthesis}}, pp. \bibinfo{pages}{8--24},
  \doi{https://doi.org/10.1109/AISPAS.1997.581620}.
\newblock \urlprefix\url{https://ieeexplore.ieee.org/document/581620}.

\bibitemdeclare{article}{Head87}
\bibitem{Head87}
\bibinfo{author}{T.~\surnamestart Head\surnameend} (\bibinfo{year}{1987}):
  \emph{\bibinfo{title}{Formal language theory and DNA: an analysis of the
  generative capacity of specific recombinant behaviours.}}
\newblock {\sl \bibinfo{journal}{Bulletin of Mathematical Biology}}
  \bibinfo{volume}{49}, pp. \bibinfo{pages}{737--759},
  \doi{https://doi.org/10.1007/BF02481771}.
\newblock \urlprefix\url{https://link.springer.com/article/10.1007/BF02481771}.

\bibitemdeclare{article}{Head-Handbook}
\bibitem{Head-Handbook}
\bibinfo{author}{T.~\surnamestart Head\surnameend},
  \bibinfo{author}{G.~\surnamestart P\u{a}un\surnameend} \&
  \bibinfo{author}{D.~\surnamestart Pixton\surnameend} (\bibinfo{year}{1997}):
  \emph{\bibinfo{title}{Language theory and molecular genetics: Generative
  mechanisms suggested by DNA recombination.}}
\newblock {\sl \bibinfo{journal}{Handbook of Formal Languages}}
  \bibinfo{volume}{2}, pp. \bibinfo{pages}{295--360},
  \doi{https://doi.org/10.1007/978-3-662-07675-0_7}.
\newblock
  \urlprefix\url{https://link.springer.com/chapter/10.1007/978-3-662-07675-0_7}.

\bibitemdeclare{book}{connection}
\bibitem{connection}
\bibinfo{author}{D.~\surnamestart Hillis\surnameend} (\bibinfo{year}{1986}):
  \emph{\bibinfo{title}{The Connection Machine.}}
\newblock \bibinfo{publisher}{MIT Press Cambridge, MA, USA},
  \doi{https://doi.org/10.1017/S026357470001002X}.

\bibitemdeclare{article}{Manea2007}
\bibitem{Manea2007}
\bibinfo{author}{F.~\surnamestart Manea\surnameend},
  \bibinfo{author}{C.~\surnamestart Mart\'{\i}n-Vide\surnameend} \&
  \bibinfo{author}{V.~\surnamestart Mitrana\surnameend} (\bibinfo{year}{2007}):
  \emph{\bibinfo{title}{Accepting networks of splicing processors: Complexity
  results.}}
\newblock {\sl \bibinfo{journal}{Theoretical Computer Science}}, pp.
  \bibinfo{pages}{72--82}, \doi{https://doi.org/10.1016/j.tcs.2006.10.015}.
\newblock
  \urlprefix\url{https://www.sciencedirect.com/science/article/pii/S0304397506007675}.

\bibitemdeclare{article}{tissue}
\bibitem{tissue}
\bibinfo{author}{C.~\surnamestart Mart\'{i}n-Vide\surnameend},
  \bibinfo{author}{J.~\surnamestart Pazos\surnameend},
  \bibinfo{author}{G.~\surnamestart P\u{a}un\surnameend} \&
  \bibinfo{author}{A.~\surnamestart Rodr\'{i}guez-Pat\'on\surnameend}
  (\bibinfo{year}{2002}): \emph{\bibinfo{title}{A new class of symbolic
  abstract neural nets: tissue P systems.}}
\newblock {\sl \bibinfo{journal}{LNCS}} \bibinfo{volume}{2387}, pp.
  \bibinfo{pages}{290--299}, \doi{https://doi.org/10.1007/3-540-45655-4_32}.
\newblock
  \urlprefix\url{https://link.springer.com/chapter/10.1007/3-540-45655-4_32}.
\newblock \bibinfo{note}{8th Annual International Conference, COCOON 2002}.

\bibitemdeclare{book}{Morrison}
\bibitem{Morrison}
\bibinfo{author}{J.P. \surnamestart Morrison\surnameend}
  (\bibinfo{year}{2010}): \emph{\bibinfo{title}{Flow-Based Programming: A New
  Approach to Application Development}}.
\newblock \bibinfo{publisher}{J.P. Enterprises}.
\newblock \urlprefix\url{https://dl.acm.org/citation.cfm?id=1859470}.

\bibitemdeclare{article}{paun}
\bibitem{paun}
\bibinfo{author}{G.~\surnamestart P\u{a}un\surnameend} (\bibinfo{year}{1998}):
  \emph{\bibinfo{title}{Distributed architectures in DNA computing based on
  splicing: Limiting the size of components.}}
\newblock {\sl \bibinfo{journal}{Unconventional Models of Computation}}, pp.
  \bibinfo{pages}{323--335}.

\bibitemdeclare{book}{membrane}
\bibitem{membrane}
\bibinfo{author}{G.~\surnamestart P\u{a}un\surnameend} (\bibinfo{year}{2002}):
  \emph{\bibinfo{title}{Membrane computing. An Introduction.}}
\newblock \bibinfo{publisher}{Springer, Berlin, Heidelberg.}

\bibitemdeclare{article}{tvd}
\bibitem{tvd}
\bibinfo{author}{G.~\surnamestart P\u{a}un\surnameend} (\bibinfo{year}{2005}):
  \emph{\bibinfo{title}{DNA computing; Distributed splicing systems.}}
\newblock {\sl \bibinfo{journal}{LNCS}} \bibinfo{volume}{1261}, pp.
  \bibinfo{pages}{353--370}, \doi{https://doi.org/10.1007/3-540-63246-8_22}.
\newblock
  \urlprefix\url{https://link.springer.com/chapter/10.10072F3-540-63246-8_22}.

\bibitemdeclare{book}{handbookNC}
\bibitem{handbookNC}
\bibinfo{author}{G.~\surnamestart Rozenberg\surnameend},
  \bibinfo{author}{T.~\surnamestart Back\surnameend} \& \bibinfo{author}{J.~Kok
  \surnamestart (Eds.)\surnameend} (\bibinfo{year}{2012}):
  \emph{\bibinfo{title}{Handbook of Natural Computing.}}
\newblock \bibinfo{publisher}{Springer, Berlin, Heidelberg.}
\newblock \urlprefix\url{https://www.springer.com/gp/book/9783540929093}.

\bibitemdeclare{book}{handbook}
\bibitem{handbook}
\bibinfo{author}{G.~\surnamestart Rozenberg\surnameend} \&
  \bibinfo{author}{A.~Salomaa \surnamestart (Eds.)\surnameend}
  (\bibinfo{year}{1997}): \emph{\bibinfo{title}{Handbook of Formal Languages.}}
\newblock \bibinfo{publisher}{Springer, Berlin, Heidelberg.}
\newblock \urlprefix\url{https://www.springer.com/gp/book/9783642638633}.

\end{thebibliography}
\end{document}